\documentclass{iopart}

\usepackage[]{graphicx}  
\usepackage{cite}

\newcommand{\E}  {\epsilon}
\newcommand{\om} {\omega}

\newcommand{\ts} {{\tilde{s}}}
\newcommand{\tu} {{\tilde{u}}}
\newcommand{\tv} {{\tilde{v}}}
\newcommand{\tw} {{\tilde{w}}}

\newcommand{\bfE}{\bi{E}}
\newcommand{\bfD}{\bi{D}}
\newcommand{\bfd}{\bi{d}}
\newcommand{\bfr}{\bi{r}}

\newcommand{\cF} {\mathcal{F}}

\begin{document}
\title{Dynamical Screening of an Atom Confined within a Finite-Width Fullerene}

\author{S Lo$^1$, A V Korol$^{1,2}$ and A V Solov'yov$^1$\footnote{On leave from the Ioffe Physical-Technical Institute, Russian Academy of Sciences, Polytechnicheskaya 26, St.~Petersburg 194021, Russia}}

\address{$^1$ Frankfurt Institute for Advanced Studies, Johann Wolfgang Goethe-Universit\"at, Max-von-Laue-Str. 1, 60438 Frankfurt am Main, Germany}
\address{$^2$ Department of Physics, St.Petersburg State Maritime Technical University, Leninskii prospect 101, St.~Petersburg 198262, Russia}

\eads{lo@fias.uni-frankfurt.de}

\begin{abstract}
This is an investigation on the dynamical screening of an atom confined within a fullerene of finite width. The two surfaces of the fullerene lead to the presence of two surface plasmon eigenmodes. It is shown that, in the vicinity of these two frequencies, there is a large enhancement of the confined atom's photoabsorption rate. 
\end{abstract}


\submitto{\jpb}

\section{Introduction}
\label{intro}
Currently there is much interest in plasmon excitations and their effects on the properties of the object (see \cite{Ilia,review,scattering,Ostling,dynamical,physreva} and references therein). Plasmons are the collective excitation of electrons. They are found in strongly correlated systems such as nanotubes, fullerenes (not limited to the carbon variety) and metallic clusters. The endohedral atom, an atom confined within a fullerene, is another area of interest \cite{dynamical,Amusia1,Amusia2,Dolmatov} and particularly metallofullerenes \cite{metallo0,metallo1,metallo2,metallo3,metallo4}. Other forms of confinement include confinement as dopant in metals \cite{confinedatoms} and confinement within hollow structures such as nanotubes and inorganic fullerenes.

Connerade and Solov'yov studied the effect of electromagnetic screening on an atom confined within a fullerene \cite{dynamical}. An atom inside the spherical $C_{60}$ molecule was considered. They found that the screening is strongly dependent on the frequency, hence the term dynamical screening. Near the plasmon frequency of the fullerene there is a large amplification of the field within the fullerene, leading to an enhancement of the photoabsorption cross section of the endohedral atom.
In their work they treated the fullerene as an infinitely thin spherical shell. Connerade and Solov'yov used the hydrodynamic formalism outlined in \cite{physreva}.

Recent work by Scully \etal \cite{prl} and by Reink\"{o}ster \etal \cite{res} demonstrate the existence of a second giant resonance in the single photoionisation cross section of multiply ionised $C_{60}$ molecules. The first resonance is known to correspond to the surface plasmon of the fullerene. The nature of the second resonance has been discussed by Scully \etal in the original paper and also by Korol and Solov'yov in the comment \cite{comment}.

This is an investigation into the effects of a fullerene that has finite width, and in particular, focusses on the dynamical screening factor studied in \cite{dynamical}.

\section{Theoretical Framework}
\label{sec:2}
\begin{figure}
\begin{center}
\includegraphics[scale=0.4]{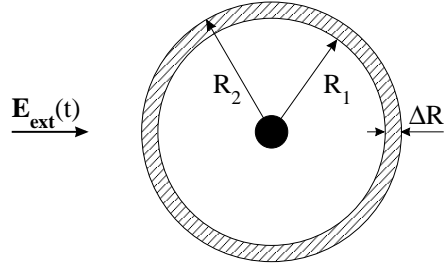}
\end{center}
\caption{The atom, indicated by the solid circle, is confined at the centre on the fullerene molecule of average radius $R$ and thickness $\Delta R$. The inner and the outer radii of the fullerene are given by $R_1$ and $R_2$ respectively. This system is exposed to an external field $\bfE_{\rm{ext}}$.}
\label{fig:1}
\end{figure}

Consider a system where an atom is confined at the centre of a spherical fullerene. This fullerene has an average radius of $R$ and a thickness of $\Delta R$ as depicted in Figure \ref{fig:1}. When this system is exposed to an electromagnetic field, the presence of the fullerene dynamically screens the confined atom - it experiences a field that is either enhanced or suppressed depending on the field frequency $\omega$. The result is that, for the same external field, the photoabsorption rate of the confined atom differs from that of the free atom. A dynamical screening factor $\cF \equiv \cF(\omega)$ may be defined to relate the photoabsorption cross sectons of these two atoms:

\begin{equation}
\label{eq:definition}
\sigma_{\rm{conf}} = \cF \sigma_{\rm{free}}.
\end{equation}

In this work the problem has been studied using a dielectric shell model, first applied to fullerenes by Lambin \etal \cite{Lambin} to study the two surface plasmon modes.
The fullerene is modelled as a spherical shell of dielectric function $\epsilon$. It has an inner and an outer radius of $R_1$ and $R_2$ respectively, with $R$ being their average. The atom, located at $r=0$ is characterised by the polarizability $\alpha_{\rm{a}}(\omega)$ and the average radius $R_{\rm{a}}$, which is assumed to be small in comparison to $R_{1,2}$. This system is exposed to a monochromatic electromagnetic wave. Using the dipole approximation and neglecting the magnetic part, this is treated as an external uniform electric field $\bfE_{\rm{ext}}(t) = \bfE_0 e^{i\omega t}$.

It is assumed that the quantities discussed here depend on the frequency $\omega$ of the external field, i.e. all the relations are written for the Fourier components of the corresponding quantities.

To calculate the screening factor $\cF(\omega)$, one must find the absorption rate of the confined atom and compare it with that of the free atom. In the dipole limit, an object exposed to an electromagnetic wave will absorb its energy with a rate Q given by \cite{Landau}

\begin{equation}
\label{eq:Q}
Q = \frac{\omega}{2} \rm{Im} \int \left( \bfE^*(\bfr) \, d\bfD(\bfr) \right).
\end{equation}


\noindent Here the integration is carried out over the volume of the considered system. $\bfE(\bfr)$ is the electric field intensity at the position $\bfr$ and $d\bfD(\bfr)$ is the dipole moment of the integration element at that position. Thus to calculate $Q_a$, the atom's photoabsorption rate, it is necessary to obtain a precise expression for the field experienced by the confined atom.

\subsection{Calculating Electric Field at the Atom}
\label{sec:thefield}

When the system is exposed to the external electric field, both the fullerene and the atom become polarized. The field, which acts on the atom may be decomposed into the contributions described by the following. The external field polarizes the dielectric shell, creating an additional field. Both fields polarize the confined atom, giving rise to a third field. This third field causes further polarization of the shell, which further polarizes the atom etc. To obtain a precise expression for the field experienced by the confined atom, one must consider this infinite cycle of the shell (and of the atom) being polarized and polarizing.

This was achieved by splitting the problem up into two parts: the dielectric shell in the external field (section \ref{sec:prob1}), and a point-like dipole at the centre of the shell in the absence of the external field (section \ref{sec:prob2}). Results from these two problems were then combined in an iterative manner to construct the final expression for  the total field $\bfE_{\rm{tot}}$ inside the fullerene (section \ref{sec:prob3}).

\subsection{Dielectric Shell in External Field}
\label{sec:prob1}

The problem of a solid dielectric sphere in a uniform field is well known and discussed in many textbooks, \cite{Landau,Jackson}. The case of a dielectric shell in a uniform field is an extension of that problem to a spherical layer \cite{fin,Anderson}. Using the method detailed in \cite{Jackson}, one finds the electric field to be:

\begin{eqnarray}
\bfE(\bfr)
=
\cases{
\displaystyle{
\bfE_0 - s\, {r^2\, \bfE_0 - 3\bfr\, (\bfE_0\bfr) \over r^5}}
& $r>R_2$
\cr
\displaystyle{
u \bfE_0 - v \, {r^2\, \bfE_0 - 3\bfr\, (\bfE_0\bfr) \over r^5}}
& $R_1< r <R_2$
\cr
\displaystyle{
w\bfE_0}
& $r<R_1$,
}
\label{eq:field1}
\end{eqnarray}

\noindent where 

\begin{eqnarray}
s =
\displaystyle{
{(\E-1)(2\E+1)\,R_2^3 + (2\E+1)(1 - \E)\,R_1^3 \over \Delta}},
\cr
u  =
\displaystyle{{3\,(2\E+1) \over \Delta}},
\quad
v  = 
\displaystyle{{3\,(1-\E)\, R_1^3  \over \Delta}},
\qquad
w =
\displaystyle{{9\,\E \over \Delta}},
\label{eq:contants1}
\end{eqnarray}

\noindent with

\begin{equation}
\label{eq:delta}
\Delta
=
(2+\E)(2\E+1)
-
2\,(1-\E)^2\,\xi^3
\end{equation}

\noindent and $\xi$, the ratio of the inner to outer shell radii, given by

\begin{equation}
\label{eq:xi}
\xi = \frac{R_1}{R_2} \leq 1.
\end{equation}

\subsection{Polarized Atom Inside Shell}
\label{sec:prob2}

When a field $\bfE$ polarizes the atom, with polarizability $\alpha_{\rm{a}}(\omega)$, the induced dipole moment $\bfd$ is given by:

\begin{equation}
\label{eq:d}
\bfd = \alpha_{\rm{a}}(\omega) \, \bfE.
\end{equation}

Consider the general problem of a point-like dipole $\bfd$ located at the centre of the dielectric shell in the absence of an external field. The dipole moment polarizes the shell. Both fields contribute to the total field intensity over the whole space. Applying the same approach as in section \ref{sec:prob1}, one finds the following expression for the electric field:

\begin{eqnarray}
\bfE(\bfr) =
\cases{
\displaystyle{
-\ts \, {r^2 \bfd - 3\bfr\,(\bfd\bfr) \over r^5} }
& $r>R_2$
\cr
\displaystyle{
\tu \,\bfd - \tv \, {r^2 \bfd - 3\bfr\,(\bfd\bfr) \over r^5} }
& $R_1<r<R_2$
\cr
\displaystyle{
\tw \, \bfd  - {r^2 \bfd - 3\bfr\,(\bfd\bfr) \over r^5} }
& $r<R_1$,
}
\label{eq:field2}
\end{eqnarray}

\noindent where 

\begin{eqnarray}
\displaystyle{
\ts = { 9\E \over \Delta}},
\qquad
\displaystyle{
\tu = { 6(1-\E) \over  R_2^3\Delta}},
\quad
\displaystyle{
\tv = { 3\,(2+\E) \over \Delta}},
\cr
\displaystyle{
\tw = {2\over R_1^3}\, {(2+\E)(\E - 1) + (1-\E) ( 2  + \E)\,\xi^3
\over \Delta}},
\label{eq:constant2b}
\end{eqnarray}

\noindent with $\Delta$ defined as in (\ref{eq:delta}).

\subsection{Total Field at the Endohedral Atom}
\label{sec:prob3}
Let us now consider the problem of the atom confined at the centre of the dielectric shell in the presence of the external field $\bfE_0$. To construct the precise expression for the field we combine the results from the previous sections in an iterative manner described below.

Let $\bfE_i$ refer to the electric field at $r=0$ due to the polarization of the shell and $\bfd_i$ refer to the induced moment of the atom in the $i^{th}$ iteration. Then the iteration scheme is as follows.

\textit{$1^{st}$ iteration:}
The external field $\bfE_0$ induces a dipole moment in the shell, which creates the additional field $\bfE_1$. The combined field $\bfE_0 + \bfE_1 = w \bfE_0$, refer to (\ref{eq:field1}), polarizes the atom and induces the dipole moment $\bfd_1 = \alpha_{\rm{a}} w \bfE_0$.

\textit{$2^{nd}$ iteration:}
As discussed in section \ref{sec:prob2}, the dipole $\bfd_1$ gives rise to another induced dipole moment of the shell. This creates the additional field $\bfE_2$ at $r=0$, refer to (\ref{eq:field2}):
\begin{equation}
\label{eq:E2}
\bfE_2 = \tw \bfd_1 = (\alpha_{\rm{a}} \tw) w \bfE_0.
\end{equation}

\noindent This field in turn induces the dipole moment $\bfd_2$ given by:

\begin{equation}
\label{eq:d2}
\bfd_2 = \alpha_{\rm{a}} \bfE_2 = (\alpha_{\rm{a}} \tw) w \alpha_{\rm{a}} \bfE_0.
\end{equation}

\textit{$i^{th}$ iteration:}
It follows that for the iterations where $i > 2$, the relations for $\bfE_i$ and $\bfd_i$ are given by:

\begin{eqnarray}
\label{eq:cyclea}
\bfE_i = \tw \bfd_{i-1},
\quad
\bfd_i = \alpha_{\rm{a}} \bfE_i.
\end{eqnarray}

\noindent The above relations are then expressed in terms of $\bfE_0$ and then summed over all iterations to give the total field due to the polarization of the system and the total dipole moment $\bfd_{\rm{tot}}$ of the atom. The total electrical field at the atom $\bfE_{\rm{tot}}$, is simply the polarization field combined with the external field.  

\begin{equation}
\label{eq:finalfield}
\bfE_{\rm{tot}} = F \bfE_0,
\qquad
\bfd_{\rm{tot}} = \alpha_{\rm{a}} F \bfE_0,
\end{equation}

\noindent where

\begin{equation}
\label{eq:Fexact}
F = w \left(1 + (\alpha_{\rm{a}}\tw) + (\alpha_{\rm{a}}\tw)^2 + \cdots \right) = \frac{w}{1 - \alpha_{\rm{a}}\tw}.
\end{equation}

The field at $r=0$, given by $\bfE_{\rm{tot}}$, has been modified by the system. Equation (\ref{eq:finalfield}) shows that the electric field, which the atom is exposed to, differs from the external field by a factor of $F
 = F(\omega)$, which depends on the frequency of the external field.

\subsection{Dynamical Screening Factor}
\label{sec:rate}

The absorption rate of the electromagnetic wave by the system may be divided into two parts: absorption due to the two surfaces of the fullerene shell and due to the atom. From (\ref{eq:Q}) and (\ref{eq:finalfield}), the atom's absorption rate is:

\begin{equation}
\label{eq:Qaexact}
Q_{\rm{a}} = \frac{\omega}{2} \rm{Im} \left(\bfE_{\rm{tot}}^*(0) \bfd_{\rm{tot}}\right)
	 = \frac{\omega}{2} \rm{Im} \left(\alpha_{\rm{a}} |F|^2 |\bfE_0|^2\right).
\end{equation}

\noindent The difference between this absorption rate for the confined atom and that of the free atom is the factor of $|F|^2$, and from (\ref{eq:definition}), this factor is the dynamical screening factor $\cF = |F|^2$. 

In the approximation that the radius of the atom, $R_{\rm{a}}$ is much smaller than that of the fullerene the expression for $\cF$ may be simplified. The polarizability of the atom, $\alpha_{\rm{a}}$, is roughly proportional to $R_{\rm{a}}^3$ \cite{Jackson}. The term $\alpha_{\rm{a}} \, \tw$ is therefore negligible and (\ref{eq:Fexact}) becomes $F \approx w$. Thus the screening factor is:

\begin{equation} 
\label{eq:cF}
\cF \approx |w|^2 = \left|\frac{9\epsilon}{\Delta}\right|^2.
\end{equation}

Further analysis of $\cF$ requires the dielectric function of the shell to be expressed in terms of the angular frequency of the electric field, $\omega$. This depends on the material nature of the fullerene. It must be emphasised that until this point the expression for the screening factor is independent of the `type of material' of the shell. Henceforth the fullerene shall be treated as being metallic on account of the delocalization of the valence electrons. Following the Drude model \cite{Bohren} and temporarily ignoring the damping of the plasmon oscillations, the dielectric function $\epsilon$ is expressed as:

\begin{equation}
\label{eq:drude}
\epsilon(\omega) = 1 - \frac{\om_{\rm{p}}^2}{\omega^2},
\end{equation}

\noindent where $\om_{\rm{p}}$ is the volume plasmon frequency, which is described by the total number of delocalised electrons $N$ (4 valence electrons from each carbon atom) and the volume of the shell $V = 4\pi(R_2^3-R_1^3)/3$:

\begin{equation}
\label{eq:wp}
\om^2_{\rm{p}} = \frac{4\pi N}{V}.
\end{equation}

It can be shown that $\Delta$, see (\ref{eq:delta}), is transformed by the Drude model into:

\begin{equation}
\label{eq:transformeddelta}
\Delta = \frac{9}{\omega^4} \left( \omega^2 - \omega_1^2 \right) \left( \omega^2 - \omega_2^2 \right )
\end{equation}

where

\begin{eqnarray}
\label{eq:frequency}
\displaystyle{\omega_1^2 = {\om_{\rm{p}}^2\over 6}\Bigl(3-p\Bigr)},
\quad
\displaystyle{\omega_2^2 = {\om_{\rm{p}}^2\over 6}\Bigl(3+p\Bigr)},
\qquad
p= \sqrt{1+8\xi^3}.
\end{eqnarray}

\noindent The frequencies $\omega_1$ and $\omega_2$ are the eigenfrequencies of the surface plasmon modes of the fullerene \cite{Ostling}: the plasmons at the inner and the outer surfaces of the fullerene are coupled oscillators and has two normal modes. In the symmetric mode characterised by $\omega_1$ the charge densities of the two surfaces oscillate in phase, whereas in the antisymmetric mode of frequency $\omega_2$ they are in antiphase.

The screening factor (\ref{eq:cF}) can be expressed as a sum of terms resonant in $\omega_1$ and in $\omega_2$:

\begin{equation}
\label{factor}
\cF(\omega) = \left| 1 + 
{2N\over R_2^3}
\left(
\frac{1}{\om^2 - \om_1^2 + \rm{i} \Gamma_{1} \om}
-
\frac{1}{\om^2 - \om_2^2 + \rm{i} \Gamma_{2} \om} 
\right)
\right|^2.
\end{equation}

\noindent 
In the expression above damping has been introduced via the widths of the symmetric and antisymmetric plasmon mode, $\Gamma_1$ and $\Gamma_2$ respectively. They describe the decay rate from collective excitation to the incoherent sum of single-electron excitations. In this paper we do not attempt to calculate these widths. Various methods for calculating the widths are described in \cite{rev24,rev85,rev87,rev88}.

\section{Results and Discussion}
\label{sec:results}

For the purposes of calculation, the widths of the two plasmons were set as a fraction of the corresponding resonant frequency: $\Gamma_{1,2} = \gamma\,\omega_{1,2}$. The values of $\gamma$ used here are 0.1, 0.25 and 0.5. It should be noted that the screening factor is now completely parameterised by the the number of delocalised electrons, the outer radii and the quantity $\xi$, the ratio of the fullerene's inner and outer radii. 

\begin{figure}
\includegraphics[scale=0.24]{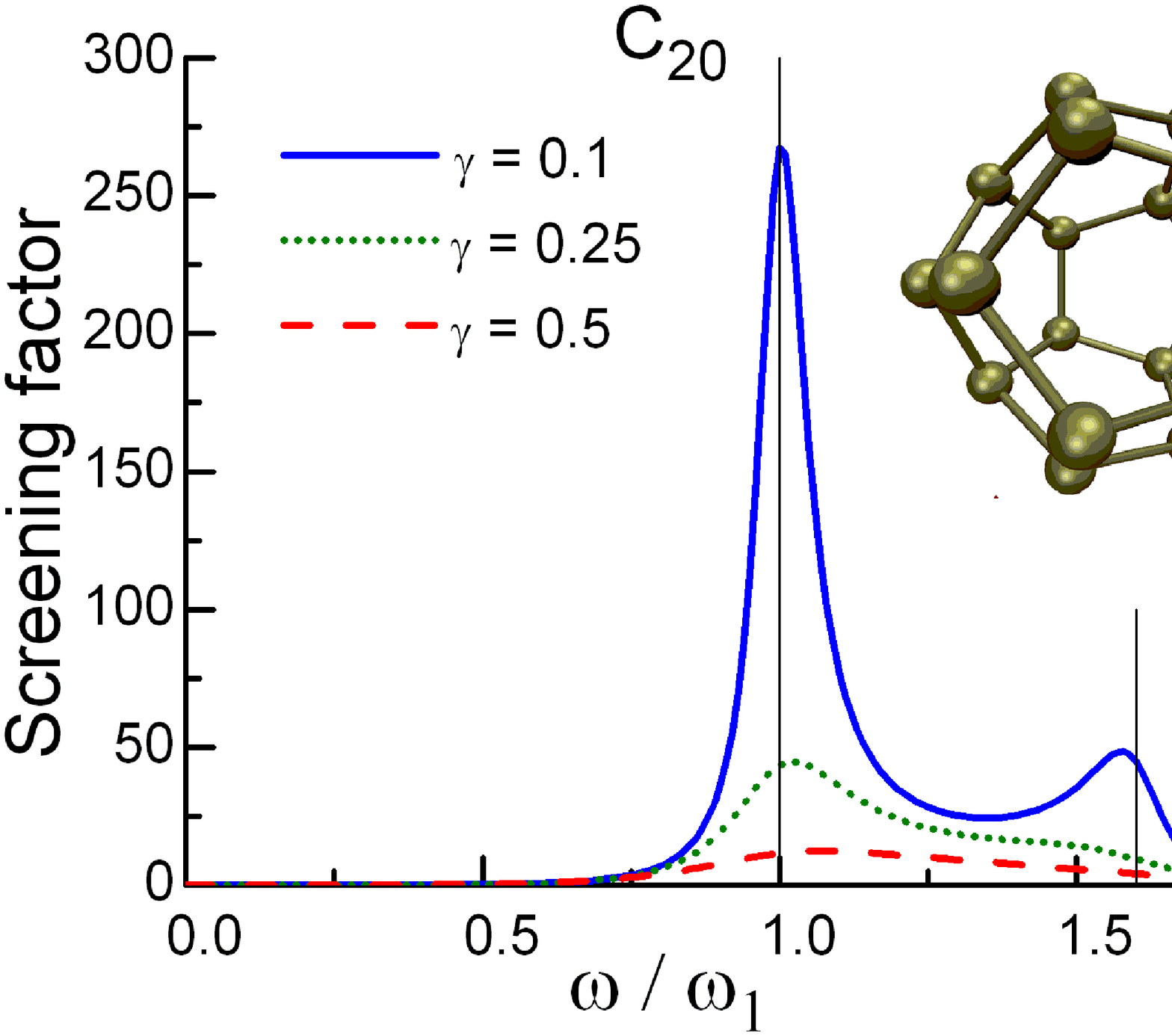}
\includegraphics[scale=0.24]{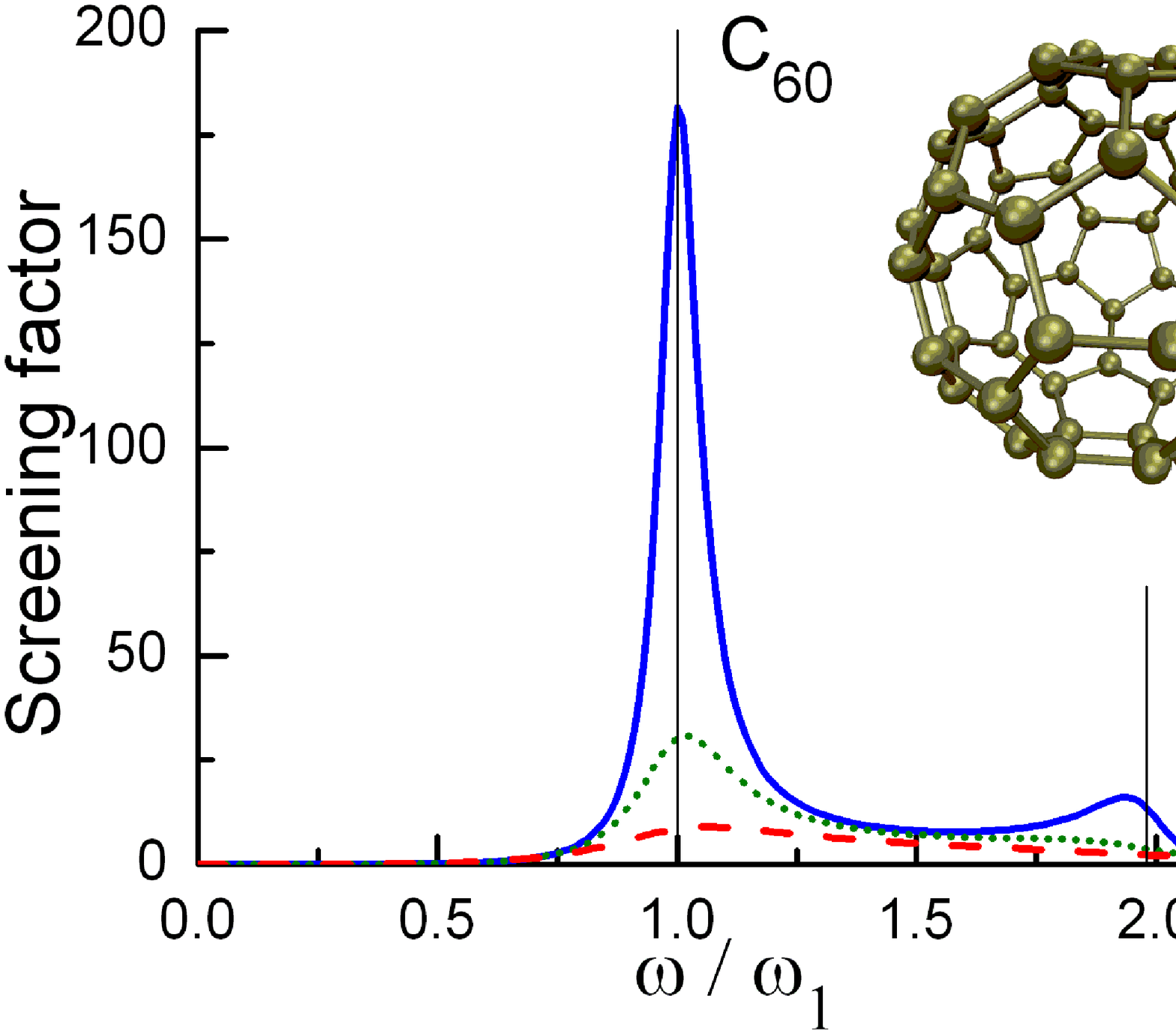}
\\
\includegraphics[scale=0.24]{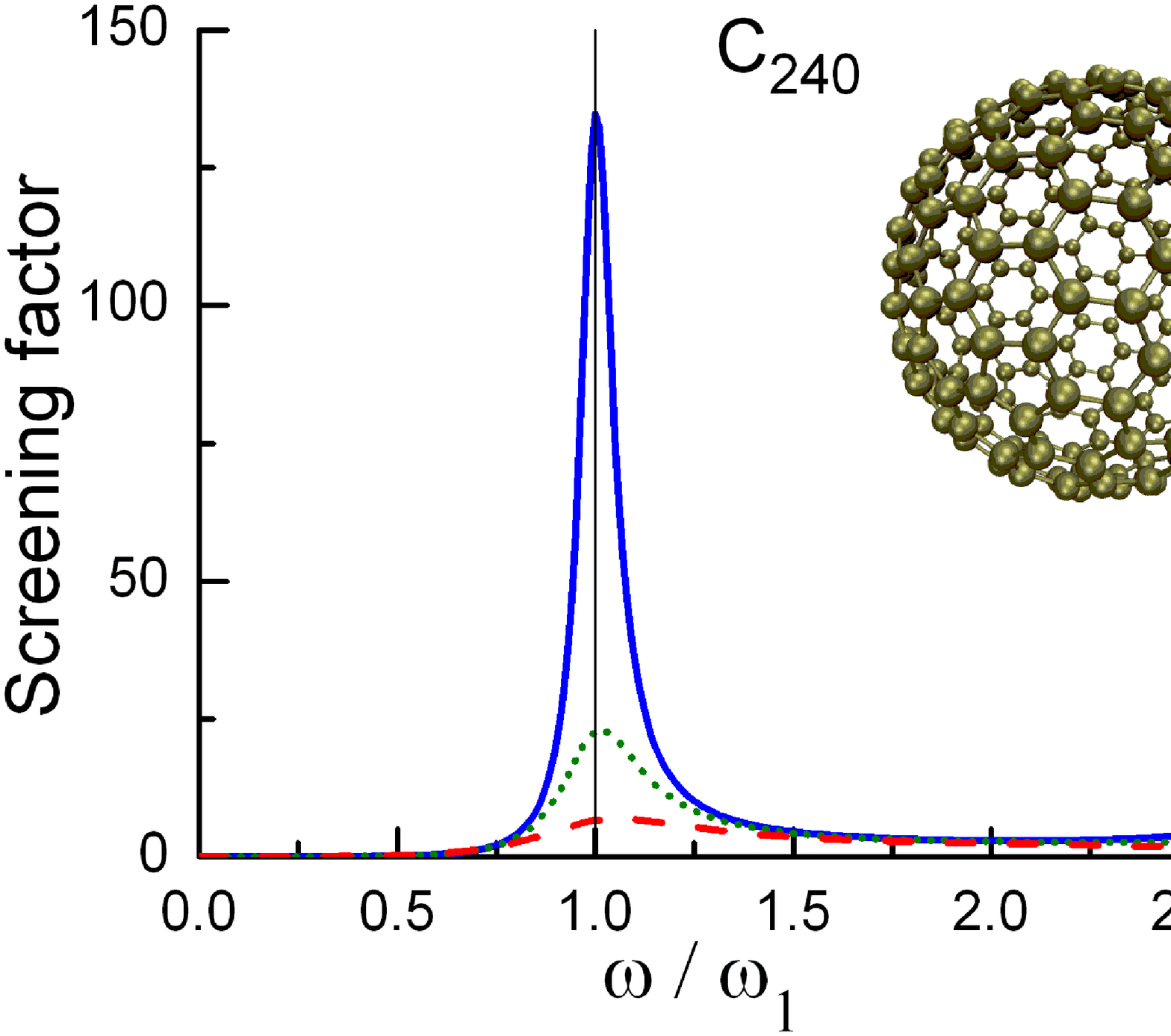}
\includegraphics[scale=0.24]{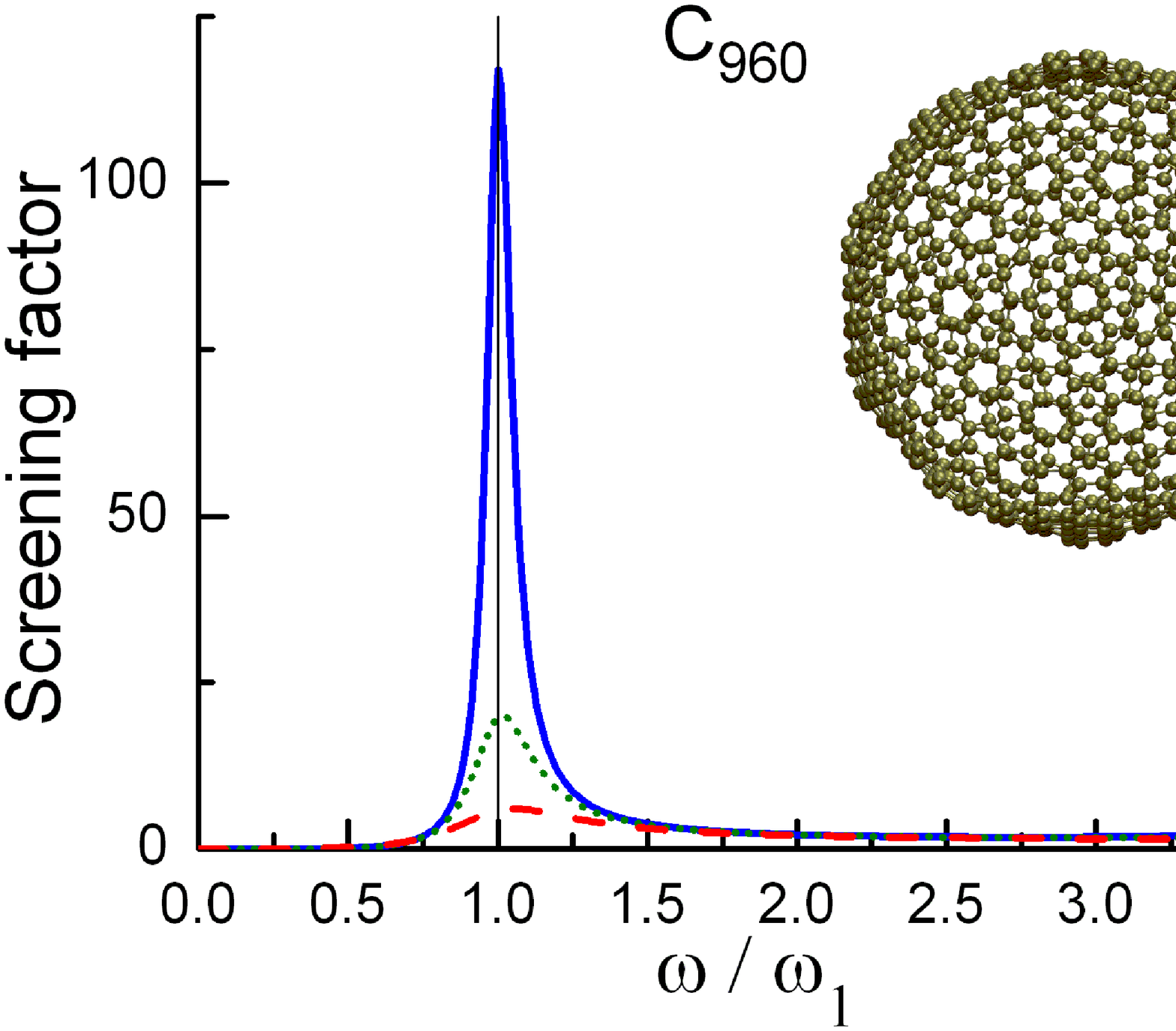}
\caption{Plots of the screening factor for the 4 fullerenes. For each fullerene, the screening factor was plotted for the same set of $\gamma$. The \textit{x}-axis is the photon energy scaled with respect to that of the first resonance. The pairs of thin vertical lines indicate the position of the two resonant frequencies for each fullerene.}
\label{fig:2}
\end{figure}

The dynamical screening factor was studied for several spherical fullerenes: $C_{20}$, $C_{60}$, $C_{240}$, and $C_{960}$. The thickness of each fullerene $\Delta R$ was set to 1.5 \AA, which was obtained by R\"udel \etal \cite{c60}. The average radii of the fullerenes are given in Table \ref{tab:1} together with other relevant properties. The plasmon frequencies are calculated from (\ref{eq:frequency}). It must be noted that $\omega_1$ for $C_{60}$ is experimentally known to be 20 eV \cite{prl}. The discrepancy between the calculated values and the known experimental values may be explained by the fact that the model is purely classical.

\begin{table}
\caption{Properties of the 4 fullerenes are given in the table: the average radius, the ratio of the inner and outer radii, and the resonant frequencies calculated from (\ref{eq:frequency}).}
\label{tab:1}
\lineup
\begin{indented}
\item[]\begin{tabular}{@{}lllll}
\br
Fullerene & R (\AA) & $\xi$ & $\omega_1$ (eV) & $\omega_2$ (eV)\\
\mr
$C_{20}$ & \02.0 \cite{c20}   & 0.45 & 19.8 & 31.7 \\
$C_{60}$ & \03.5 \cite{c60}   & 0.65 & 16.9 & 33.5 \\
$C_{240}$ & \07.1 \cite{radii}& 0.81 & 12.8 & 35.0 \\
$C_{960}$ & 13.8 \cite{radii} & 0.89 & \09.9  & 37.1 \\
\br
\end{tabular}
\end{indented}
\end{table}

The resulting plots of the screening factor for the four fullerenes are given in Figure \ref{fig:2}. The $x$-axis is the photon energy scaled with respect to $\omega_1$ to aid comparison between the different fullerenes. It should be noted that $C_{20}$ has the largest $\omega_1$.

\begin{figure}
\begin{center}
\includegraphics[scale=0.23]{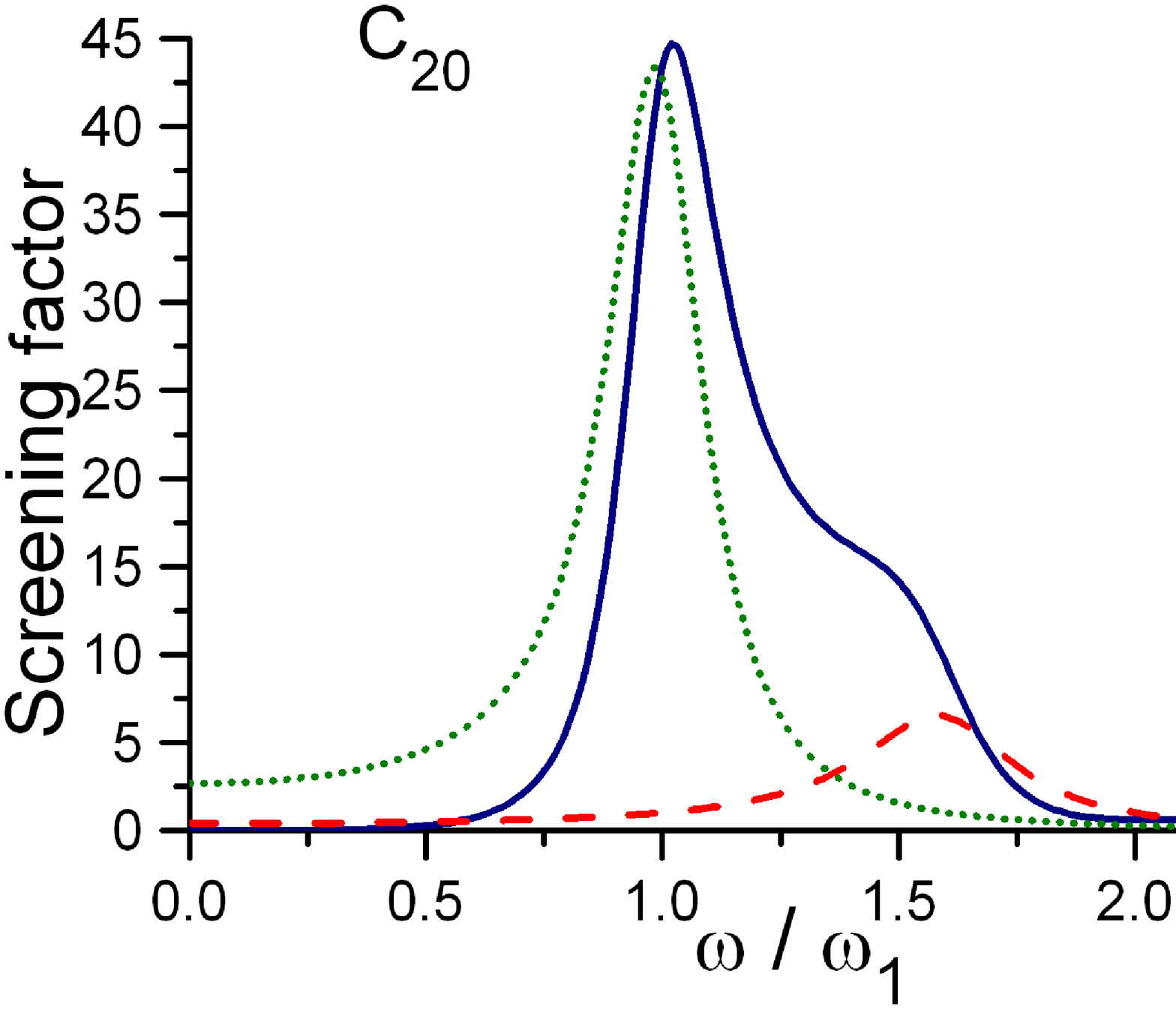}
\includegraphics[scale=0.23]{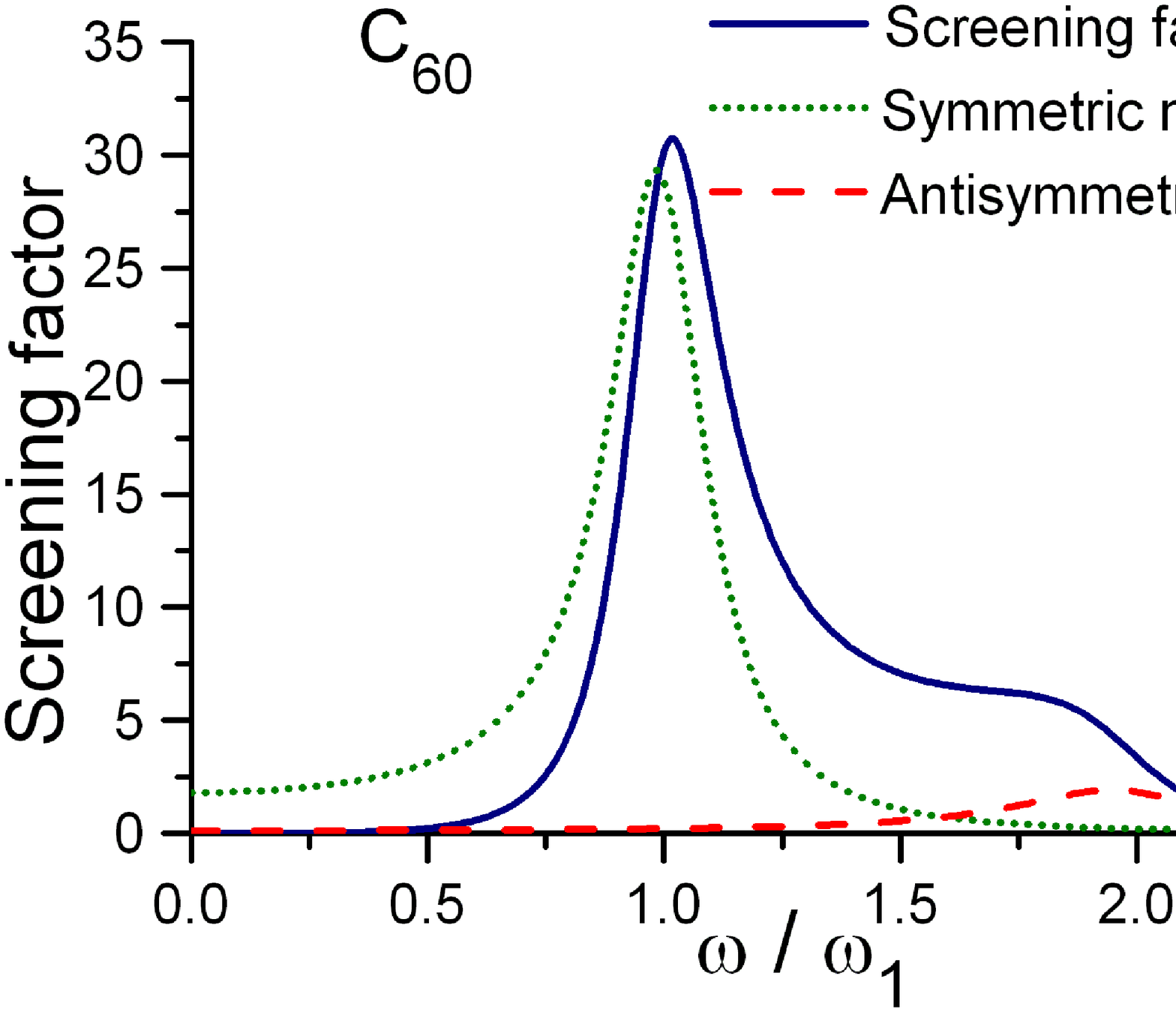}
\end{center}
\caption{A plot of the screening factor for $C_{20}$ (left panel) and $C_{60}$ (right panel) with $\gamma = 0.25$. The individual contributions of the symmetric and the antisymmetric plasmon modes are also shown.}
\label{fig:3}
\end{figure}

All the plots of the screening factor show that in the static limit the atom is completely shielded from the external field. At every high frequencies the fullerene is transparent to the light and there is no enhancement. This is as expected for a metallic fullerene. The profile of the screening factor for the different fullerenes are very similar and show a strong dependence on the value of $\gamma$. There is a peak in the vicinity of the first resonant frequency and another feature near the second resonant frequency due to the symmetric and the antisymmetric plasmon modes, respectively, as demonstrated in Figure \ref{fig:3}. Their characteristics are determined by the damping of the plasmons, which are determined by the value of $\gamma$. For each fullerene, there is a well defined peak at each resonant frequency for $\gamma = 0.1$. For larger values of $\gamma$, the individual contribution of the second plasmon mode is strongly suppressed by the damping. Here the interference terms of the screening factor become more significant and the second resonance is reduced to an extension of the first peak, as shown in Figure \ref{fig:3}.

From Figure \ref{fig:2} it can be seen that the relative heights of the first peaks are almost the same for the different fullerenes. In this region the contribution from the second plasmon mode is effectively constant and is small. Using the approximation that the maximum occurs at the first resonant frequency, one arrives at this result:

\begin{equation}
\label{eq:fmax}
\cF_{\rm{max}} \approx \cF(\omega_1)
\approx \left( \frac{2N}{pR_2^3}\frac{1}{\gamma \omega_1^2} \right)^2
= \left( \frac{3+p}{2p} \frac{1}{\gamma} \right)^2,
\end{equation}

\noindent where $p$ is a function of $\xi$, which is fixed by the particular fullerene. This relation explains the large differences in $\cF_{\rm{max}}$, seen in Figure \ref{fig:2}, for the different values of $\gamma$ and also between the different fullerenes.

\subsection{Infinitely Thin Fullerene}
\label{sec:limit}
In the limit of the infinitely thin fullerene of radius $R$, $\xi \rightarrow 1$. It is useful to note that $(1-\xi^3) = (p+3)(p-3)/8$. Following on from this, one shows that the two resonant frequencies become

\begin{eqnarray}
\label{eq:limfreq}
\cases{
\displaystyle{\lim_{\xi\to 1}\omega_1^2 
= \lim_{\xi\to 1}{\omega_{\rm{p}}^2\over 6}\Bigl(3-p\Bigr)
= \frac{2N}{3R^2}} \equiv \omega_{\rm{f}}^2
\cr
\displaystyle{\lim_{\xi\to 1}\omega_2^2 
= \lim_{\xi\to 1}{\omega_{\rm{p}}^2\over 6}\Bigl(3+p\Bigr)
\rightarrow \infty}
}
\end{eqnarray}

Physically, the two surfaces approach each other. At the limit, the two surfaces merge and only the symmetric mode is possible. Here, $\omega_2 \rightarrow \infty$, removing the second resonant term, and $\omega_1$ becomes $\omega_{\rm{f}}$, the resonant frequency for the surface plasmon of the infinitely thin fullerene. Thus the screening factor becomes:

\begin{equation}
\label{eq:limf}
\lim_{\xi \to 1} \cF = \left| 1 + \frac{\omega_{\rm{f}}^2}{\omega^2-\omega_{\rm{f}}^2 + \rm{i}\Gamma_1\omega} \right|^2 
= \frac{\omega^4 + \Gamma_1^2 \omega^2 } {(\omega^2 - \omega_{\rm{f}}^2)^2 + \Gamma_1^2\omega^2}
\end{equation}



\begin{figure}
\begin{center}
\includegraphics[scale=0.25]{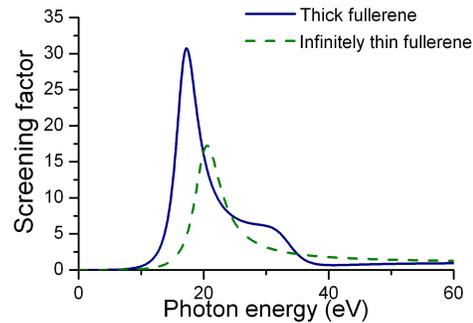}
\end{center} 
\caption{A plot comparing the screening factors for the infinitely thin fullerene \cite{dynamical} and that of the fullerene of finite thickness. This is done for $C_{60}$ with $\gamma = 0.25$.}
\label{fig:4}
\end{figure}

\noindent This is the result given by Connerade and Solov'yov in \cite{dynamical}.
A comparison of the two screening factors is given in Figure \ref{fig:4}. For $C_{60}$ $\omega_f$ is 20.2 eV.

The distinguishing feature of the screening factor of the thick fullerene is the presence of the second peak. Recent studies \cite{prl,res} have demonstrated the existence of a second plasmon peak in the photoabsorption cross section of the fullerene.

\section{Conclusion}

In this paper we have generalised the dynamical screening factor obtained by Connerade and Solov'yov \cite{dynamical} for an atom confined within an infinitely thin fullerene to that of an atom confined by a fullerene of finite width. We have taken a classical dielectrical approach to study the screening of an atom at the centre of the fullerene. We have focussed on the case where the polarisability of the confined atom is sufficiently small so as to not have an appreciable effect on the field inside the fullerene.

With this classical approach we give a qualitative description of the dynamical screening factor. We have shown that there are two surface plasmon eigenomdes of this thick fullerene. They result in two frequency ranges where there are large enhancements of the confined atom's photoabsorption cross section. This is in contrast to the single frequency range for the infinitely thin fullerene. For cases of minor damping of the plasmons, there are two distinct peaks of the screening factor in the vicinity of the plasmon resonant frequencies. When damping is large, the second peak is suppressed and becomes an extension of the first peak.

Despite the simplicity of this model, it provides an instructive description of the phenomena discussed here. This model may be further applied to study the dynamic responses of more complicated situations various processes.

\ack
We are grateful for the helpful discussions with Prof. Dr. W. Greiner.

This work was supported by INTAS under the grant 03-51-6170, by the  European Commission within the Network of Excellence project EXCELL (project number 515703) and by PECU under the grant 004916(NEST).


\end{document}